\documentclass[10pt,conference]{IEEEtran}

\usepackage{amssymb}
\usepackage{amsfonts}
\usepackage{amsmath}
\usepackage{latexsym}
\usepackage{epsfig}
\usepackage{psfrag}
\usepackage{bm}
\usepackage{xspace}
\usepackage{graphicx}

\renewcommand{\baselinestretch}{0.995}

\newtheorem{claim}{Claim}
\newtheorem{proposition}[claim]{Proposition}

\newtheorem{theorem}[claim]{Theorem}
\newtheorem{definition}{Definition}

\newcommand{\eproposition}{\hfill$\square$}

\newcommand{\etheorem}{\hfill$\square$}
\newcommand{\edefinition}{\hfill$\square$}

\newcommand{\ignore}[1]{}
\newcommand{\comment}[1]{{\LARGE #1}}
\newcommand{\TG}{\hat{{\cal N}}}
\newcommand{\bb}{\hat{b}}
\newcommand{\W}{W}
\newcommand{\normal}{{\cal N}}
\newcommand{\mess}{\tau}

\newcommand{\nbr}{N}
\newcommand{\Udot}{\dot{U}}

\newcommand{\vlambda}{\boldsymbol{\lambda}}
\newcommand{\vomega}{\boldsymbol{\omega}}
\newcommand{\bvomega}{\bar{\boldsymbol{\omega}}}

\newcommand{\R}{\mathbb{R}}

\newcommand{\mutriv}{\mu_{\mathrm{triv}}}
\newcommand{\muQtwo}[1]{\mu_{\mathrm{Q}2,#1}}
\newcommand{\muT}[1]{\mu_{\mathrm{T},#1}}

\newcommand{\phiML}{\phi_{\mathrm{ML}}}
\newcommand{\phiLP}{\phi_{\mathrm{LP}}}

\newcommand{\Pmudec}[2]{P_{#1}^{#2}}

\newcommand{\PmutrivLP}{\Pmudec{\mutriv}{\phiLP}}
\newcommand{\PmuQtwoLP}[1]{\Pmudec{\muQtwo{#1}}{\phiLP}}
\newcommand{\PmuTLP}[1]{\Pmudec{\muT{#1}}{\phiLP}}

\newcommand{\matr}[1]{\mathbf{#1}}
\newcommand{\vect}[1]{\mathbf{#1}}

\newcommand{\vu}{\vect{u}}

\newcommand{\vx}{\vect{x}}
\newcommand{\bvx}{\vect{\bar x}}

\newcommand{\hbvx}{\vect{\hat {\bar x}}}

\newcommand{\tr}{\mathsf{T}}

\newcommand{\set}[1]{\mathcal{#1}}
\newcommand{\code}[1]{\mathcal{#1}}
\newcommand{\defeq}{\triangleq}

\newcommand{\bcode}[1]{\code{\bar #1}}
\newcommand{\bset}[1]{\set{\bar #1}}

\newcommand{\convhull}{\operatorname{conv}}

\newcommand{\graph}[1]{\mathsf{#1}}

\newcommand{\diamgv}[2]{\Delta_{#1}(\graph{#2})}

\newcommand{\Nmax}{N^{\mathrm{max}}}

\newcommand{\wpsAWGNC}{w_{\mathrm{p}}^{\mathrm{AWGNC}}}

\newcommand{\dv}{d_{\mathrm{v}}}
\newcommand{\dc}{d_{\mathrm{c}}}

\newcommand{\xpar}{\kappa}

\begin{document}

\title{The Benefit of Thresholding in \\ 
       LP Decoding of LDPC Codes}

\author{\authorblockN{Jon Feldman}
\authorblockA{Dept.~of Industrial Engineering\\
              and Operations Research \\
              Columbia University \\
              New York, NY 10027, USA \\
              \texttt{jonfeld@ieor.columbia.edu}}
\and
\authorblockN{Ralf Koetter}
\authorblockA{Coordinated Science Laboratory \\
              and Dept.~of ECE \\
              University of Illinois \\
              Urbana, IL 61801, USA \\
              \texttt{koetter@uiuc.edu} }
\and
\authorblockN{Pascal O.~Vontobel}
\authorblockA{Dept.~of Electrical and \\
              Computer Engineering \\
              University of Wisconsin \\
              Madison, WI 53706, USA \\
              \texttt{vontobel@ece.wisc.edu}}
}

\maketitle

\begin{abstract}
  Consider data transmission over a binary-input additive white Gaussian noise
  channel using a binary low-density parity-check code. We ask the following
  question: Given a decoder that takes log-likelihood ratios as input, does it
  help to modify the log-likelihood ratios before decoding?  If we use an
  optimal decoder then it is clear that modifying the log-likelihoods cannot
  possibly help the decoder's performance, and so the answer is ``no.''
  However, for a suboptimal decoder like the linear programming decoder, the
  answer might be ``yes'': In this paper we prove that for certain interesting
  classes of low-density parity-check codes and large enough SNRs, it is
  advantageous to truncate the log-likelihood ratios before passing them to
  the linear programming decoder.
\end{abstract}

\section{Introduction}
\label{sec:introduction:1}

While maximum-likelihood (ML) decoding of low-density parity-check (LDPC)
codes is reasonably well understood based on the expected weight distribution
of the codes,
the linear programming (LP) and the related belief propagation (BP)
decoding of LDPC codes reveal a number of interesting and unexpected
phenomena. The root cause of the
difference between these suboptimal decoders and ML decoding is the occurrence
of so called {\em pseudo-codewords}; from the perspective of an LP or BP
decoder, the pseudo-codewords act as attractive solutions to the decoding
problem, even though they are not actual codewords in the LDPC code under
consideration. In contrast to codewords which, for codes of length $n$ and
under antipodal signaling, map to elements of the set $\{+1,-1\}^n$,
pseudo-codewords are vectors of length $n$ that map to vectors with entries
that lie in the interval $[-1,+1]$. Note that the set of possible
pseudo-codewords is a function not only of the code but also of the chosen
parity-check matrix.

\begin{figure}
  \begin{center}
    \epsfig{file=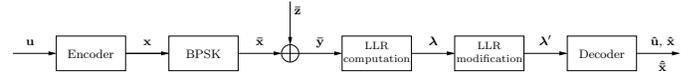, width=\linewidth}
  \end{center}
  \caption{Communication setup under consideration. (See main text for
           explanations.)}
  \label{fig:comm:setup:1:1}
\end{figure}

This paper explores one of the above-mentioned unexpected phenomena of LP
decoding and discusses the roots of this behavior.  Considering the tight
relationship between LP decoding and iterative
decoding~\cite{Koetter:Vontobel:03:1, Vontobel:Koetter:04:2, Feldman:03:1,
Feldman:Wainwright:Karger:05:1},
our observations about LP decoding must also have consequences for
iterative decoding.  Before we start describing that phenomenon, let
us first explain the communication setup (see
Fig.~\ref{fig:comm:setup:1:1}) that is under consideration.
\begin{itemize}

  \item We use a binary channel code of length $n$, dimension $k$, and rate
    $k/n$.

  \item The information word $\vu \in \{0, 1\}^k$ is encoded into the codeword
    $\vx \in \{ 0, 1 \}^n$. We assume that all information words are chosen
    with equal likelihood.

  \item Let $\theta:\ \R \to \R$, $\omega_i \mapsto 1 - 2
    \omega_i$. Restricting the domain of $\theta$ to $\{ 0, 1 \}$ we obtain
    the usual BPSK mapping: $0 \mapsto +1$ and $1 \mapsto -1$. When applying
    the map $\theta$ to a vector we define the result to be a vector where
    each component is mapped according to $\theta$. Instead of
    $\theta(\omega_i)$ and $\theta(\vomega)$ we will very often simply write
    $\bar \omega_i$ and $\bvomega$, respectively. For our communication setup
    this means that the codeword $\vx \in \{0, 1 \}^n$ is mapped to its
    signal-space point $\bvx \defeq \theta(\vx) = \big( \theta(x_1), \ldots,
    \theta(x_n) \big) \in \{ +1, -1 \}^n$.

  \item For $i = 1, \ldots, n$, the symbols $\bar x_i$ are sent over a
    (binary-input) additive white Gaussian noise channel (AWGNC) with noise
    power $N_0/2$, i.e.~we receive $\bar Y_i \defeq \bar x_i + \bar Z_i$ where
    $\{ \bar Z_i \}_{i=1}^{n}$ are i.i.d.~random variables with $\bar Z_i \sim
    \normal(0,N_0/2)$. Here, $\normal(\mu,\sigma^2)$ denotes a Gaussian random
    variable with mean $\mu$ and variance $\sigma^2$.

  \item Based on the observations
$\bar Y_i = \bar y_i$, $i = 1, \ldots, n$, we
    compute the normalized log-likelihood ratios (LLRs)
    {\small
    \begin{align*}
      \lambda_i
        &\defeq
           \eta
           \cdot
           \log
             \left(
               \frac{p_{\bar Y_i|\bar X_i}(\bar y_i| +\!1 )}
                    {p_{\bar Y_i|\bar X_i}(\bar y_i| -\!1 )}
             \right)
         = \eta
           \cdot
           \log
             \left(
               \frac{p_{\bar Y_i|X_i}(\bar y_i|0)}
                    {p_{\bar Y_i|X_i}(\bar y_i|1)}
             \right),
    \end{align*}
    }%
    where the normalization constant $\eta \defeq \eta(N_0)$ is chosen such
    that $\lambda_i$ equals $+1$ if $\bar z_i = 0$.
    
  \item A mapping $\mu: \R \to \R$ is applied to the LLRs and results in the
    modified LLRs $\lambda'_i \defeq \mu(\lambda_i)$, $i = 1, \ldots, n$.

  \item Based on the modified LLR vector $\vlambda'$, a decoder $\phi$ tries
    to make a decision $\hbvx \defeq \phi(\vlambda')$ about $\bvx$. (Or,
    alternatively, tries to decide on $\vu$ or $\vx$.)

  \item When decoding a code of length $n$, we use the label
    $\Pmudec{\mu}{\phi}(n)$ for denoting the block error probability of a
    decoder $\phi$ which bases its decisions on the modified LLR vector
    $\vlambda' \defeq \mu(\vlambda)$.

\end{itemize}

Let $\bcode{C} \defeq \theta(\code{C})$ be the set of points in signal space
that correspond to the codewords. Using the (normalized) LLR vector
$\vlambda$, the maximum likelihood (ML) decoder $\phiML$ can be cast as
\begin{align}
  \hbvx
    &\defeq 
       \phiML(\vlambda')
     \defeq
       \arg \max_{\bvx \in \bcode{C}}
         \sum_{i=1}^{n}
           \bar x_i \lambda'_i,
             \label{eq:ml:decoder:1}
\end{align}
with the trivial mapping $\lambda'_i \defeq \mutriv(\lambda_i) \defeq
\lambda_i$, $i = 1, \ldots, n$. From this expression it is clear the the LLR
vector $\vlambda$ is a sufficient statistic for optimal decoding. Moreover,
using the data-processing inequality (see e.g.~\cite{Cover:Thomas:91}) it can
easily be shown that there is no mapping $\mu$ such that for a given code of
length $n$ there is a decoder $\phi$ such that $\Pmudec{\mu}{\phi}(n) <
\Pmudec{\mutriv}{\phiML}(n)$.

The situation is not as simple in the case of suboptimal decoders, e.g.~the
linear programming (LP) decoder~\cite{Feldman:03:1,
Feldman:Wainwright:Karger:05:1}. In fact, combining the results in~\cite{Feldman:Malkin:Stein:Servedio:Wainwright:04:1}
and~\cite{Koetter:Vontobel:03:1},
we show that for certain
low-density parity-check (LDPC) codes and for high enough SNR it is favorable
\emph{not} to use the trivial map $\mutriv$, but to use a 
two-level quantization map
\begin{align*}
  \lambda'_i
    &\defeq
       \muQtwo{L}(\lambda_i)
     \defeq
       \begin{cases}
         +L & \text{if $\lambda_i \geq 0$}  \\
         -L & \text{if $\lambda_i < 0$}
       \end{cases}
\end{align*}
before performing the LP decoding.  

This seeming paradox is not uncommon for
suboptimal algorithms. We cite the following paragraph from Ganti et
al.~\cite[p.~2316]{Ganti:Lapidoth:Telatar:00:1} which remarks on a similar
phenomenon (albeit in a different context): ``[\ldots] Indeed, in the matched
case it is clear that the optimal decoder for the general channel performs at
least as well as a decoder that first quantizes the output and then performs
optimal processing on the quantized samples. Under mismatched decoding,
however, it is unclear how to relate the performance of the mismatched decoder
on the original channel to its performance on the output-quantized channel.''

A natural question arises: Is the 
advantage of using the two-level quantization map 
the result of a quantization effect, or something else?
We show that
there are code families such that for any finite $W$, the thresholding map
\begin{align}
\label{eq:muT}
  \lambda'_i
    &\defeq
       \muT{W}(\lambda_i)
     \defeq
       \begin{cases}
         +W        & \text{if $\lambda_i \geq +W$}  \\
         -W        & \text{if $\lambda_i \leq -W$} \\
         \lambda_i & \text{otherwise}
       \end{cases}
\end{align}
is also favorable to
the trivial map $\mutriv$.  This suggests that the asymptotic
advantage over $\mutriv$ is gained not by quantization, but rather by
restricting the LLRs to have finite support.

The rest of the paper is structured as follows. We will give a brief
introduction to LP decoding and pseudo-codewords in
Sec.~\ref{sec:lp:decoding:1}.\footnote{For recent work on the notion of
pseudo-codewords in decoding we refer to
\cite{Koetter:Li:Vontobel:Walker:04:1, Vontobel:Koetter:04:1,
Vontobel:Koetter:04:2, Koetter:Vontobel:03:1, Feldman:Stein:05:1,
Feldman:03:1, Feldman:Wainwright:Karger:05:1}. } In Sec.~\ref{sec:canonical:completion:1}, we will talk about
pseudo-codewords stemming from the canonical completion and their importance
for the asymptotic behavior of the LP decoder. In
Secs.~\ref{sec:quantizing:thresholding:1} and~\ref{sec:proof}, we will discuss the main results of
this paper, namely we show examples when thresholding and quantizing of the
LLRs can help.

\section{LP Decoding}
\label{sec:lp:decoding:1}

ML decoding as in \eqref{eq:ml:decoder:1} can also be formulated as
\begin{align}
  \hbvx
    &\defeq
       \phiML(\vlambda')
     \defeq
       \arg \max_{\bvx \in \convhull(\bcode{C})}
         \sum_{i=1}^{n}
           \bar x_i \lambda'_i,
             \label{eq:ml:decoder:2}
\end{align}
where $\convhull(\bcode{C})$ is the convex hull of $\bcode{C}$ and where the
mapping $\mu$ is the trivial mapping $\mutriv$. Unfortunately, for most codes
of interest, the description complexity of $\convhull(\bcode{C})$ grows
exponentially in the block length and therefore finding the maximum in
\eqref{eq:ml:decoder:2} with a linear programming solver is highly impractical
for reasonably long codes.\footnote{Exceptions to this observation include for
example the class of convolutional codes with not too many states.}

A standard approach in optimization in order to simplify the problem, is to
replace the maximization over $\convhull(\bcode{C})$ by a maximization over
some easily describable polytope $\bset{P}$ that is a relaxation of
$\convhull(\bcode{C})$:
\begin{align}
  \hbvx
    &\defeq \arg \max_{\bvx \in \bset{P}}
       \sum_{i=1}^{n}
         \bar x_i \lambda'_i.
           \label{eq:lp:decoder:1}
\end{align}
If $\bset{P}$ is strictly larger than $\convhull(\bcode{C})$ then the decision
rule in \eqref{eq:lp:decoder:1} obviously represents a sub-optimal decoder. A
relaxation which works particularly well for LDPC codes is given by the
following approach~\cite{Feldman:03:1, Feldman:Wainwright:Karger:05:1}. Let
$\code{C}$ be described by an $m \times n$ parity-check matrix $\matr{H}$ with
rows $\vect{h}_1, \vect{h}_2, \ldots, \vect{h}_m$. Then the polytopes $\set{P}
\defeq \set{P}(\matr{H})$ and $\bset{P} \defeq \bset{P}(\matr{H}) \defeq
\theta(\set{P})$, also called the fundamental
polytopes~\cite{Koetter:Vontobel:03:1}, are defined as

\vspace{-0.3cm}
{\small
\begin{align*}
  \set{P}
    &\defeq
       \bigcap_{i=1}^{m}
         \convhull(\code{C}_i)
  \text{ with }
  \code{C}_i
     \defeq \big\{
              \vx \in \{0, 1\}^n
              \, | \, 
              \vect{h}_i \vx^\tr = 0
              \, \operatorname{mod}\, 2
            \big\}, \\
  \bset{P}
    &\defeq
       \bigcap_{i=1}^{m}
         \convhull(\bcode{C}_i)
  \text{ with }
  \bcode{C}_i
     \defeq \theta(\code{C}_i).
\end{align*}
}%
Note that $\set{P}$ is a convex set within $[ 0, 1 ]^n$ that contains
$\convhull(\code{C})$ but whose description complexity is much smaller than
the description complexity of $\convhull(\code{C})$. (A similar comment
applies to $\bset{P}$ which is a convex set within $[-1,+1]^n$ and which
contains $\convhull(\bcode{C})$.) Points in the set $\set{P}$ will be called
pseudo-codewords, and since $\set{P}$ is a convex polytope, we may restrict
our attention to the vertices of $\set{P}$ (and $\bset{P}$).  Because the set
$\bset{P}$ is usually strictly larger than $\convhull(\bcode{C})$, the
decoding rule in~\eqref{eq:lp:decoder:1} might deliver a vertex of $\bset{P}$
that is not the signal-space equivalent of a codeword; these ``fractional''
vertices are the reason for the sub-optimality of LP decoding
(cf.~\cite{Feldman:Wainwright:Karger:05:1,Koetter:Vontobel:03:1}).

For analyzing the above setup it turns out to be useful to define the AWGNC
pseudo-weight~\cite{Forney:Koetter:Kschischang:Reznik:01:1} of a
pseudo-codeword $\vomega \in \set{P}$ to be $\wpsAWGNC(\vomega) =
||\vomega||_1^2 / ||\vomega||_2^2$, where $||\vomega||_1$ and $||\vomega||_2$
are the $L_1$- and $L_2$-norm of $\vomega$, respectively. The significance of
$\wpsAWGNC(\vomega)$ is the following. The existence of a pseudo-codeword
$\vomega = (\omega_1, \omega_2, \ldots, \omega_n) \in \set{P} \setminus \{
\vect{0} \}$ causes LP decoding to fail to detect the codeword $\vect{0}$ if
the vector of received LLRs $\vlambda = (\lambda_1, \lambda_2, \ldots,
\lambda_n)$ satisfies the inequality $\sum_{i=1}^{n} \bar \omega_i \cdot
\lambda'_i > \sum_{i=1}^{n} \bar 0 \cdot \lambda'_i$, where $\vlambda' =
\mutriv(\vlambda) = \vlambda$. Then it can be shown that the squared Euclidean
distance from $\vect{\bar 0} = +\vect{1}$ to the plane $\big\{ \vlambda' \in
\R^n \ | \ \sum_{i=1}^{n} (\bar \omega_i - \bar 0) \lambda'_i = 0 \big\}$ is
$\wpsAWGNC(\vomega)$.

\section{The Canonical Completion and \\ 
         its Implications}
\label{sec:canonical:completion:1}

Consider a $(\dv,\dc)$-regular\footnote{An LDPC code is called a
$(\dv,\dc)$-regular code if the uniform column weight of the relevant
parity-check matrix $\matr{H}$ is $\dv$ and the uniform row weight of
$\matr{H}$ is $\dc$.}  binary code $\code{C}$ of length $n$ described by a
parity-check matrix $\matr{H}$. Its Tanner graph~\cite{Tanner:81} will be
denoted by $\graph{T} \defeq \graph{T}(\matr{H})$, where the set of variable
nodes will be called $V \defeq V(\graph{T})$, the set of check nodes will be
called $C \defeq C(\graph{T})$, and a node $v \in V$ is adjacent to a node $c
\in C$ if and only if the corresponding entry in $\matr{H}$ equals $1$. Given
a variable node $v \in V$, we let $\diamgv{v}{T}$ denote the maximal (graph)
distance from $v$ that any node in $\graph{T}$ can have. Our goal in this
section is to construct a pseudo-codeword whose impact on the LP decoder
depends on the mapping $\mu$. Before defining this pseudo-codeword, we need a
definition.

\begin{definition}[\!\!\cite{Koetter:Vontobel:03:1}]
  \label{def:tanner:graph:ordering:1}

  Let $\graph{T}$ be a Tanner graph. We denote an arbitrary variable node $v
  \in V(\graph{T})$ to be the root. We classify the remaining variable and
  check nodes according to their (graph) distance from the root, i.e.~the root
  is at tier 0, all nodes at distance $1$ from the root will be called nodes
  of tier $1$, all nodes at distance $2$ from the root node will be called
  nodes of tier $2$, etc.. We call this ordering ``breadth-first spanning tree
  ordering with root $v$.'' Because of the bipartiteness of $\graph{T}$, it
  follows easily that the nodes of the even tiers are variable nodes whereas
  the nodes of the odd tiers are check nodes. Furthermore, a check node at
  tier $2t+1$ can only be connected to variable nodes in tier $2t$ and
  possibly to variable nodes in tier $2t+2$. Note that the last tier is tier
  $\diamgv{v}{T}$ and that the variable nodes are at tiers $0, 2, \ldots, 2
  \lfloor \diamgv{v}{T}/2 \rfloor$. \edefinition
\end{definition}

\begin{definition}[Canonical completion~\cite{Koetter:Vontobel:03:1}]
  \label{def:canonical:completion:1}

  Let $\code{C}$ be a binary $(\dv,\dc)$-regular code with parity-check matrix
  $\matr{H}$ and Tanner graph $\graph{T} \defeq \graph{T}(\matr{H})$. Let $v
  \in \graph{T}$ be an arbitrary variable node. After performing the
  breadth-first spanning tree ordering with root $v$, we construct a vector
  $\tilde \vomega$ in the following way. If bit $i$ corresponds to a variable
  node in tier $2t$, then
  \begin{align*}
    \tilde \omega_i \defeq \frac{1}{(\dc-1)^t}.
  \end{align*}
  It is possible to choose a scaling factor $\alpha > 0$ (in fact, a whole
  interval of $\alpha$'s) such that $\vomega \defeq \alpha \cdot \tilde
  \vomega \in \set{P}(\matr{H})$. We call the resulting pseudo-codeword
  $\vomega$ the canonical completion with root $v$. \edefinition
\end{definition}

\begin{theorem}[\!\!\cite{Koetter:Vontobel:03:1}]
  \label{theorem:min:ps:weight:canonical:completion:1}

  Same scenario as in Def.~\ref{def:canonical:completion:1}. The canonical
  completion with root $v$ yields a vector $\vomega$ such that $\vomega$ is in
  the fundamental polytope $\set{P}(\matr{H})$. Imposing the additional mild
  constraint $3 \leq \dv < \dc$, the pseudo-weight $\wpsAWGNC(\vomega)$ of
  $\vomega$ can be upper bounded by
  \begin{align*}
    \wpsAWGNC(\vomega)
      &\leq
         \beta'_{\dv,\dc} \cdot n^{\beta_{\dv,\dc}},
  \end{align*}  
  where {\small
  \begin{align*}
    \beta'_{\dv,\dc}
      &\defeq \left( 
                \frac{\dv (\dv-1)}
                     {\dv-2}
              \right)^2,
    \
    \beta_{\dv,\dc}
      \defeq \frac{\log\left( (\dv-1)^2 \right)}
                  {\log\big( (\dv-1)(\dc-1) \big)}
      <      1.
  \end{align*}
  }
  \etheorem
\end{theorem}

Assuming $\mu$ to be the trivial mapping $\mutriv$, the above theorem has
immediate consequences for the LP decoder: the LP decision region for
$\vect{\bar 0}$ is constrained by a hyperplane whose squared Euclidean
distance from $\vect{\bar 0}$ is at most $\beta'_{\dv,\dc}
n^{\beta_{\dv,\dc}}$. Because $\beta_{\dv,\dc} < 1$, this implies that the
word error probability $\PmutrivLP(n)$ of LP decoding is {\em lower} bounded:
$\PmutrivLP(n) \geq \big(1-1/(K'n^{\beta_{\dv,\dc}}) \big) \big( 2\pi K'
n^{\beta_{\dv,\dc}} \big)^{-1/2} \exp \big( -\frac{K'}{2} n^{\beta_{\dv,\dc}}
\big)$ where $K'$ is positive and a function of the SNR,
independent of $n$.  This observation
implies that the reliability function $\lim_{n \rightarrow \infty} \sup
-\frac{1}{n} \log \big( \PmutrivLP(n) \big)$ of the AWGNC under LP decoding
approaches zero for any fixed SNR.  This is in stark contrast to ML decoding
whose reliability function remains non-zero for large enough signal-to-noise
ratios. In this context it is interesting to note that Lentmaier et
al.~\cite{Lentmaier:Truhachev:Costello:Zigangirov:04:1} could prove that under
some mild technical conditions the block error rate of a $(\dv,\dc)$-regular
code under belief-propagation decoding with a bounded number of iterations is
{\em upper} bounded by $P_{\mathrm{tree}}(n) \leq n \cdot \exp(-K''
n^{\beta_{\dv,\dc}/4})$ for the \emph{same} constant $\beta_{\dv,\dc}$, where
$P_{\mathrm{tree}}(n)$ refers to the block error rate of a belief propagation
decoding algorithm where the number of iterations is one quarter the girth of
the Tanner graph.

\section{Quantizing and Thresholding}
\label{sec:quantizing:thresholding:1}

We still consider the LP decoder, but we want to investigate what happens when
$\mu$ is selected to be something other than $\mutriv$. So, let us consider
what happens when $\mu \defeq \muQtwo{L}$ is selected for some\footnote{Note
that the result of the LP decoder is independent of the exact choice of $L >
0$.} $L >0$. Actually, it can easily be seen that the combination of the AWGNC
and this quantization gives (apart from scaling) the same LLR vectors as at
the receiver end of a binary symmetric channel (BSC). Recognizing this, we can
use the results of~\cite{Feldman:Malkin:Stein:Servedio:Wainwright:04:1} which
show that there exists families of expander-based $(\dv,\dc)$-regular LDPC
codes which are guaranteed to correct a constant fraction $\tau$ of errors on
the BSC. By a simple union bound argument we conclude that for sufficiently
large SNR the block error probability is upper bounded by $\PmuQtwoLP{L}(n)
\leq n \exp(-K''' n)$
where again $K'''$ is positive and independent of $n$.
It follows that there exist families of expander-based $(\dv,\dc)$-regular
LDPC codes where $\lim_{n \rightarrow \infty} \sup -\frac{1}{n}\log \big(
\PmuQtwoLP{L}(n) \big)$ is strictly larger than zero under LP
decoding, for sufficiently large SNR.  

What explains this advantage in the asymptotic behavior?  Looking at the above
results we have to consider two candidates: \textbf{(i)} the quantized values
of the modified LLRs or \textbf{(ii)} the finite support of the modified
LLRs. It turns out that the answer is given by (ii), namely it is sufficient
to threshold the LLRs, whereas quantization as in (i) is not really
necessary. As is shown in the Section~\ref{sec:proof}, one can set $\mu \defeq
\muT{W}$ (see~\eqref{eq:muT}) for any finite $W \geq 1$ and construct classes of
$(\dv,\dc)$-regular expander-based LDPC codes where $\lim_{n \rightarrow
\infty} \sup -\frac{1}{n}\log \big( \PmuTLP{W}(n) \big)$ is non-zero under LP
decoding.\footnote{The constraint $W \geq 1$ is not necessary, but was imposed
to simplify the presentation; Th.~\ref{th:main:1} holds for any $W > 0$.}

\begin{theorem}
  \label{th:main:1}

  Consider the setup as described in Sec.~\ref{sec:introduction:1} where we
  transmit over an AWGNC with noise power $\sigma^2 \defeq N_0/2$. For any
  finite truncation value $W \geq 1$, any constant rate $0 < r < 1$, and
  sufficiently small $\sigma^2 > 0$, there exists a family of
  $(\dv,\dc)$-regular Tanner graphs for low-density parity-check codes of
  increasing length, each with rate at least $r$, such that $\lim_{n
  \rightarrow \infty} \sup -\frac{1}{n} \log \big( \PmuTLP{W}(n) \big)$ is
  strictly larger than zero.
\end{theorem}

\begin{proof}
  See Section~\ref{sec:proof}.
\end{proof}

Putting the above results for the LP decoding with the different mappings $\mu
= \mutriv$ and $\mu = \muT{W}$ in juxtaposition reveals a surprising property
of LP decoding. For values of SNR where both the lower bound on $\PmutrivLP$
and the upper bound on $\PmuTLP{W}$ are non-trivial it is actually
advantageous for (certain classes of) long codes to threshold the LLRs before
attempting to decode. In other words, since there is an $n$ large enough (as a
function of $K$ and $K'''$) such that $n \exp(-K''' n)$ is less than
$(1-1/(K'n^{\beta_{\dv,\dc}})) (2\pi K' n^{\beta_{\dv,\dc}})^{-1/2}
\exp(-\frac{K'}{2} n^{\beta_{\dv,\dc}})$, operating on the thresholded
versions of the LLRs
will
yield a smaller probability of error than retaining the full information
contained in $\vlambda$.\footnote{A similar comment can be made about LP
decoding with $\mu = \mutriv$ vs.~$\mu = \muQtwo{L}$: there is an $n$ from
where on it is better to work with the one-bit quantized LLRs than with the
original LLRs.}

What does this mean for a pseudo-codeword $\vomega$ associated with a canonical
completion? Roughly speaking, the mappings $\muT{W}$ and $\muQtwo{L}$ bend the
vector $\vlambda$ in such a way that the pseudo-codeword $\vomega$ is less
often the result of the LP decoder. This bending, which for an optimal decoder
can only deteriorate its performance, turns out to be overall helpful for a
sub-optimal algorithm like the LP decoder, at least for certain interesting
classes of LDPC codes and large enough SNRs.

\section{Proof of Theorem~\ref{th:main:1}}
\label{sec:proof}

This Section is devoted to proving Th.~\ref{th:main:1}. Before we start going
through the different steps of the proof, we introduce some useful
notation. For an integer $n$, we use $[n]$ to denote the set of integers from
$1$ to $n$. We use $\graph{T}(n,m)$ to denote a Tanner graph with $n$ variable
nodes and $m$ check nodes. For such a Tanner graph, we will usually identify
the set of variable nodes $V$ with $[n]$ and the set of check nodes $C$ with
$[m]$. For a set of nodes $S$, let $\nbr(S)$ denote
the neighbor set of $S$.

\begin{definition} 
  A Tanner graph $\graph{T}$ with variable node set $V$ of size $n$, is an
  {\em $(\alpha n, \beta)$-expander} if all sets $S \subseteq V$ with $|S|
  \leq \alpha n$ have $|\nbr(S)| \geq \beta |S|$. \edefinition
\end{definition}

The following proposition follows from~\cite{Burshtein:Miller:01} (see
also~\cite{Spielman:95:1}):

\begin{proposition} 
  \label{prop:expand:1}
  Let $0 < r < 1$, and let $\dv$ and $\dc$ be positive integers such that
$r = 1 - \frac{\dv}{\dc}$.  Then for any $0 < \delta < 1 -
\frac{1}{\dc}$, and sufficiently large $n$, there exists a Tanner
graph with $n$ variable nodes, $m = n \dv / \dc$ check nodes, uniform
variable node degree $\dv$, and uniform check degree $\dc$, which is
an $(\alpha n, \delta \dv)$-expander, where $0 < \alpha < 1$ is a
constant that does not depend on $n$. Moreover, a randomly constructed
graph has these properties with high probability.\eproposition
\end{proposition} 

For the given truncation value $W$ in Th.~\ref{th:main:1}, let $\dv$
be any integer greater than $4(4W+2)$.  Let $\hat{\delta}$
be any constant where $1 - \frac 1 \dv > \hat{\delta} > 1 -
\frac{3}{4} (\frac 1 {4W+2})$.  Now let $\delta$ be the largest value
that is less than or equal to $\hat{\delta}$ such that $\delta \dv$ is
an integer.  Note that $\hat{\delta} - \delta \leq \frac 1 \dv$.  This
implies that $\delta > 1 - \frac{1}{4W+2}$.

From Prop.~\ref{prop:expand:1}, we obtain a family of Tanner graphs;
each graph $\graph{T}(n,m)$ has uniform variable degree $\dv$, uniform
check degree $\dc$, has $r = 1 - {\frac m n }$, and is an $(\alpha n,
\delta \dv)$-expander, for some constant $\alpha$ that does not depend
on $n$.  Fix a particular length $n$, and call $\code{C} \defeq
\code{C}(n,m)$ the code defined by the Tanner graph $\graph{T} \defeq
\graph{T}(n,m)$ from the family.

Suppose the vector $+\vect{1} = \vect{\bar 0} \in \bcode{C}$ is transmitted
over the AWGNC.  Define $U \defeq \{i \in [n] : \lambda'_i < 1/2\}$, where
$\lambda'$ is defined according to~\eqref{eq:muT}.\footnote{The value 1/2 in
the definition of $U$ was set for simplicity.  The main theorem will go
through for any $W > 0$, as long as this constant ``1/2'' is less than 1,
greater than zero, and less than or equal to $W$.} This set represents the
variable nodes with ``high noise.''  For one particular $i \in [n]$, define
$p(\sigma^2)$ as the probability that $i \in U$.  Note that $p(\sigma^2)$ is
the same for all $i$, is a function only of the variance $\sigma^2$, and goes
to zero as $\sigma^2$ goes to zero.  

Define $\gamma \defeq \frac{(1-\delta)\dv}{(1 - \delta)\dv + 1}$.  Note that $0
< \gamma < 1$.  
Let $\sigma^2$ be sufficiently small so that 
$p(\sigma^2) < \frac{\alpha}{2(1 + \gamma)}$.
  By a simple Chernoff
bound 
we have
that
\begin{align}
  |U|
    &\leq \frac{\alpha n}
               {2(1 + \gamma)}
     \leq \frac{\alpha n - 1}
               {1 + \gamma}
  \label{eq:bound:1}
\end{align}
with probability at least $1 - 2^{-\Omega(n)}$.  In other words, with high
probability, the set of nodes with high noise is ``small.''

We let $\delta' \defeq 2\delta - 1$ and define
\begin{align*}
  \Udot
    &\defeq
       \bigg\{
          \bigg.
            i \in V 
          \ \bigg| \
            i \notin U \text{ and } 
            | \nbr(i) \cap \nbr(U) | > (1 - \delta') \dv
       \bigg\}.
\end{align*}
The set $\Udot$ represents the variable nodes that do not have high noise, but
do have high connectivity to the neighbors of the nodes with high noise.

We appeal to the following, which uses the same argument as a 
similar theorem in~\cite{Feldman:Malkin:Stein:Servedio:Wainwright:04:1}:

\begin{theorem}
  If $\graph{T}$ is an $(\alpha n, \delta \dv)$-expander and $|U| \leq
  \frac{\alpha n - 1}{1 + \gamma}$ then $|U| + |\Udot| \leq \alpha
  n$. \etheorem
\end{theorem}

Using \eqref{eq:bound:1} together with this theorem, we have that $|U| +
|\Udot| \leq \alpha n$ with probability at least $1 - 2^{-\Omega(n)}$.  At
this point we will apply what we know about the expansion of the graph to
prove that the LP decoder succeeds. We first need another definition and
proposition from~\cite{Feldman:Malkin:Stein:Servedio:Wainwright:04:1}:

\begin{definition}[\!\!\cite{Feldman:Malkin:Stein:Servedio:Wainwright:04:1}]
  A $\delta$-{\em matching} of $U$ is a subset $M$ of the edges incident to
  $U' \defeq U \cup \Udot$ such that {\bf (i)} every check node incident to at
  most one edge of $M$, {\bf (ii)} every node in $U$ is incident to at least
  $\delta \dv$ edges of $M$, and {\bf (iii)} every node in $\Udot$ is incident
  to at least $\delta' \dv$ edges of $M$. \edefinition
\end{definition}

\begin{proposition}[\!\!\cite{Feldman:Malkin:Stein:Servedio:Wainwright:04:1}]
  If $\graph{T}$ is an $(\alpha n, \delta \dv)$-expander with $\delta
  \dv$ an integer, and $|U| + |\Udot| \leq \alpha n$, then $U$ has a
  $\delta$-matching. \eproposition
\end{proposition}

It remains to show how the existence of a $\delta$-matching proves that the LP
decoder will succeed.  To prove that the LP decoder succeeds, we use the
method of finding a {\em dual witness}.  More details, as well as a general
treatment of this technique, can be found
in~\cite{Feldman:Malkin:Stein:Servedio:Wainwright:04:1, Feldman:Stein:05:1}.
Here, we state the definition and theorem relevant to this application:

\begin{definition}[\!\!\cite{Feldman:Malkin:Stein:Servedio:Wainwright:04:1}]
  \label{def:weights}

  Given a Tanner graph $\graph{T}(n,m)$, and a vector of LLRs $\lambda'_i$, a
  setting of weights $\{ \mess_{ij} \}$ to the edges $(i,j)$ in $\graph{T}$ is
  {\em feasible} if {\bf(i)} for all checks $j \in [m]$ and distinct $i,i' \in
  \nbr(j)$, we have $\mess_{ij} + \mess_{i'j} \geq 0$, and {\bf (ii)} for all
  nodes $i \in [n]$, we have $\sum_{j \in \nbr(i)} \mess_{ij} <
  \lambda'_i$. \edefinition
\end{definition}

\begin{theorem}[\!\!\cite{Feldman:Malkin:Stein:Servedio:Wainwright:04:1}]
  \label{thm:LPsuccess}

  Under any memoryless binary-input output-symmetric channel,
  using any binary linear code, under the assumption that $+\vect{1} =
  \vect{\bar 0}$ is transmitted, the LP decoder (using a Tanner graph
  $\graph{T}$ for the code) succeeds if and only if there exists a feasible
  weight assignment to the edges of $\graph{T}$.

\etheorem
\end{theorem}

Finally, using a line of reasoning similar
to~\cite{Feldman:Malkin:Stein:Servedio:Wainwright:04:1}, we establish that a
$\delta$-matching is sufficient to guarantee a feasible edge weight
assignment, and thus a proof that the LP decoder succeeds.  Here is where we
use our bound on $\delta$ in terms of $W$:

\begin{theorem}
  If $U$ has a $\delta$-matching, and $\delta > 1 - \frac{1}{4W + 2}$, then
  there exists a feasible edge weight assignment. \etheorem
\end{theorem}

\begin{proof}
  Given a $\delta$-matching $M$, we assign weights $\mess_{ij}$ to each edge
  $(i,j)$ in the graph as follows; we later specify the parameter $\xpar > 0$.
  \begin{itemize}
  
    \item For all $j$ such that $(i,j) \in M$ for some $i \in U$, set
      $\mess_{ij} \defeq -\xpar$, and set $\mess_{i'j} \defeq \xpar$ for all $i' \in
      \nbr(j) \setminus \{ i \}$.

    \item For all other $j$, set $\mess_{i,j} \defeq 0$ for all $i \in
      \nbr(j)$.

\end{itemize}

This weighting clearly satisfies condition (i) of a feasible weight assignment.
For the second condition, there are three cases.
\begin{enumerate}

  \item For a variable node $i \in U$, we have $-W \leq \lambda'_i < 1/2$.  By
    definition of $M$, at least $\delta \dv$ edges incident to $i$ have
    $\mess_{ij} = -\xpar$. All other incident edges have $\mess_{ij} \in \{0,
    \xpar\}$, and so the total weight of edges incident to $i$ is at most
    $\delta \dv (-\xpar) + (1-\delta) \dv \xpar$ = $(1 - 2 \delta) \dv \xpar$.
    If we maintain {\bf (a)} $\xpar > \frac{W}{(2 \delta -1) \dv}$, then this
    total weight less than $-W$, which is less or equal to $\lambda'_i$, as
    required.

  \item For a variable node $i \in \Udot$, we have $\lambda'_i \geq
    1/2$.  At least $\delta' \dv$ edges incident to $i$ are in $M$,
    and therefore have weight $0$, by the definition of $M$ and the
    weight assignment.  All other edges have weight $0$ or $+\xpar$.
    Therefore the total weight of incident edges is at most $(1 -
    \delta') \dv \xpar = 2(1 - \delta) \dv \xpar$. If we maintain {\bf
    (b)} $\xpar < \frac{1}{4(1 - \delta) \dv}$, then this total weight
    is less than $1/2$, which is less or equal to $\lambda'_i$, as
    required.

\item For a variable node $i \notin (U \cup \Udot)$, by definition
this variable node has at least $\delta' \dv$ edges not incident to
$N(U)$.  These edges all have weight 0, and so we get the same
condition {\bf (b)} as in the previous case.

\end{enumerate}
Combining our requirements {\bf (a)} and {\bf (b)} on $\xpar$, we get the
overall requirement $\frac{2\delta - 1}{4(1 - \delta)} > W$, which is
equivalent to our assumption on $\delta$.
\end{proof}

Putting it all together, we have shown that for an arbitrary truncation value
$W$, and rate $r$, there is a sufficiently small $\sigma^2$ and a family of
$(\dv,\dc)$-regular graphs on which the LP decoder succeeds with probability
$1 - 2^{-\Omega(n)}$ when $+\vect{1} = \vect{\bar 0}$ is transmitted over an
AWGNC with noise power $\sigma^2$ and with LLR modification $\mu \defeq
\muT{W}$. The assumption that $+\vect{1} = \vect{\bar 0}$ is transmitted is
without loss of generality because the polytope is ``$\code{C}$-symmetric''
(see~\cite{Feldman:Wainwright:Karger:05:1, Feldman:03:1} for details). Thus we
have shown that the word error rate of the LP decoder decreases exponentially.

\section*{Acknowledgments}

J.F.'s research was supported by NSF Mathematical Sciences Postdoctoral Research Fellowship DMS-0303407.

R.K.'s research was supported by NSF Grants CCR 99-84515 and CCR 01-05719.

P.O.V.'s research was supported by NSF Grants CCR 99-84515, CCR 01-05719,
ATM-0296033, DOE SciDAC, and ONR Grant N00014-00-1-0966.

\bibliographystyle{ieeetr}
\bibliography{feldman_koetter_vontobel_isit2005_final1_arxiv1}

\end{document}